\documentclass{article}
\usepackage{spconf,amsmath,graphicx,xcolor}

\usepackage{times}
\usepackage{epsfig}
\usepackage{graphicx}
\usepackage{amsmath}
\usepackage{amssymb}

\usepackage{adjustbox}
\usepackage{gensymb}

\usepackage[breaklinks=true,bookmarks=false]{hyperref}
\hypersetup{
    colorlinks=true,
    linkcolor=blue,
    filecolor=magenta,      
    urlcolor=red,
}
\title{Domain adaptation, Explainability \& Fairness in AI for Medical Image Analysis: Diagnosis of COVID-19 based on 3-D Chest CT-scans  }

\name{Dimitrios Kollias$^1$ \qquad  Anastasios Arsenos$^2$ \qquad  Stefanos Kollias$^{2,3}$}
\address{$^1$  Queen Mary University of London, UK \\ {\tt d.kollias@qmul.ac.uk} \\
$^2$ National Technical University of Athens, Greece \\
$^3$ National Infrastructures for Research and Technology, Greece}

\begin{document}
%
\maketitle
\begin{abstract}

The paper presents the DEF-AI-MIA COV19D Competition, which is organized in the framework of the 'Domain adaptation, Explainability, Fairness in AI for Medical Image Analysis (DEF-AI-MIA)' Workshop of the 2024 Computer Vision and Pattern Recognition (CVPR) Conference. The Competition is the 4th in the series, following the first three Competitions held in the framework of  ICCV 2021, ECCV 2022 and ICASSP 2023 International Conferences respectively. It includes two Challenges on: i) Covid-19 Detection and ii) Covid-19 Domain Adaptation. The Competition use data from COV19-CT-DB database, which is described in the paper and includes a large number of  chest CT scan series. Each chest CT scan series consists of a sequence of 2-D CT slices, the number of which is between 50 and 700. Training, validation and test datasets have been extracted from COV19-CT-DB and provided to the participants in both Challenges. The paper presents the baseline models used in the Challenges and the  performance which was obtained respectively.

\end{abstract}

\begin{keywords}
deep neural networks, domain adaptation, explainability, AI, diagnosis, 4th COVID-19 Competition, COV19-CT-DB database
\end{keywords}

\section{Introduction}

In the past few years, Deep Learning (DL) techniques have made rapid advances in many medical image analysis tasks. In pathology and radiology applications, they managed to increase the accuracy and precision of medical image assessment, which is often considered subjective and not optimally reproducible. This is due to the fact that they can extract more clinically relevant information from medical images than what is possible in current routine clinical practice by human assessors. Nevertheless, considerable development and validation work lies ahead before AI-based methods can be fully integrated ad used in routine clinical tasks. 

Of major importance is research on domain adaptation, fairness and explainability in AI-enabled medical image analysis. This research constitutes the main target of the Domain adaptation, Explainability and Fairness in AI for Medical Image Analysis (DEF-AI-MIA) Workshop, to be held in the 2024 Computer Vision and Pattern Recognition (CVPR) International Conference.  The DEF-AI-MIA workshop aims to foster discussion and presentation of ideas to tackle these challenges in the field, as well as identify research opportunities in this context. It is the fourth in the AI-MIA series of Workshops, which includes the Workshops held at IEEE ICASSP 2023, ECCV 2022 and ICCV 2021 Conferences.

This Workshop’s focus is also motivated by recent actions and regulatory policies developed in Europe and considered worldwide. GRNET, the Greek National Infrastructures for Research and Technology, has implemented the integration of public hospital units in GRNET academic network, to support research and clinical activities in medicine and biology, also providing an archiving service for data produced by the imaging devices of the hospitals at the GRNET health data centers. At the European level, EU has been regulating a European Health Data Space, which: a) fosters a genuine single market for electronic health record systems, relevant medical devices and high risk AI systems (primary use of health data), b) generates a consistent, trustworthy and efficient set-up for the use of health data for research and innovation (secondary use of health data; GRNET is involved in the implementation of this set-up). The above are linked to the recent EU AI-Act regulatory framework for AI, which classifies AI systems used in different applications according to the risk they pose to users. These are under consideration, by the public and the private sector, in Europe, USA and other countries all over the world. 

Topics covered in the workshop are domain adaptation, explainability, fairness, for trustworthiness in AI-enabled medical imaging which include a digital pathology and radiology images; use of self-supervised and unsupervised methods to enforce shared patterns emerging directly from data, develop strategies to leverage few (or partial) annotations, promoting interpretability in both model development and/or results obtained, ensure generalizability to data coming from multi-centers, multi-modalities or multi-diseases, in edge, or cloud frameworks, and robustness to out of distribution data. 

Technologies and topics to be addressed in the DEF-AI-MIA Workshop include the following: explainable 2-D \& 3D-CNN, CNN-RNN, transformer, foundation models, multimodal Large Language Models, unsupervised, self-supervised Machine Learning (ML) models for medical diagnosis; sensing “salient features” of AI/ML models related to decision-making, in spatial (images), temporal (video), volumetric (3-D) data;  optimal visualization of salient features and areas in the input data; Low/Middle/High level feature extraction \& analysis for model interpretatability and explainability;   explanation of which features and at what time, or slice, or respective intervals, are the most prominent for the provided decision in temporal and 3-D data; explainable data correlations for predictions in data streams of multimodal data; joint optimization of positive and negative saliencies; global and local models for prediction or classification; attention and self-attention mechanisms in DL/AI approaches; interpretability at training time through adversarial regularization; learning new data (from multiple sources) by leveraging knowledge already extracted and codified,    through domain adaptation; generalizable ML/DL methods when the training medical image datasets are small; generalizable ML/DL methods in cases of images with potential domain shift; unsupervised, weakly supervised and semi-supervised model adaptation; uncertainty estimation and quantification, self-training; adaptation and prompt engineering in Foundation Models (e.g., LLMs) for explainable decisions and prediction; algorithmic fairness; zero/one shot learning, avoidance of catastrophic forgetting.

\section{The 4th COV19D Competition}

A variety of technologies have been developed for early diagnosis of Covid-19, based on medical image analysis, especially focusing on 3-D chest CT scans. Special interest has been given to combined segmentation and classification approaches \cite{wang2021deep}, targeting detection of abnormalities, including consolidation, ground-glass opacities,  interlobular septal lung thickening, mostly under pleura.

The 4th COV19D Competition is the 4th in the series of COV19D Competitions following the first 3 Competitions we organized in the framework of ICCV 2021 \cite{ref99}, ECCV 2022 \cite{ref100} and ICASSP 2023 \cite{kollias2023ai} Workshops respectively. It includes two Challenges: i) Covid-19 Detection Challenge and ii) Covid-19 Domain Adaptation Challenge.

Both Challenges are  based on the COV19-CT-DB database, briefly described next, including 3-D chest CT scan series. Each chest CT scan series consists of a sequence of 2-D CT slices, the number of which is between 50 and 700.

\subsection{Covid-19 Detection Challenge}

Many CT scans have been aggregated, each one of which has been manually annotated in terms of Covid-19 and non-Covid-19 categories. The resulting dataset is split into training, validation and test partitions. The training and validation sets along with their annotations have been provided to the Competition participating teams to develop AI/ML/DL models for Covid-19 and non-Covid-19 prediction. Performance of the different approaches will be evaluated on the test set in terms of the ‘macro’ F1 score.

\subsection{Covid-19 Domain Adaptation Challenge}

CT scans have been aggregated from various hospitals and medical centres. Each CT scan has been manually annotated with respect to Covid-19 and non-Covid-19 categories. The resulting dataset is split into training, validation and test partitions. Participants have been provided with a training set that consists of: i) the annotated data of the 1st Challenge which are aggregated from some hospitals and medical centres (case A); ii) a small number of annotated data and a larger number of non-annotated data (case B), all of which are aggregated from other hospitals and medical centres and their distribution is different from that of case A. Participants have been also provided with a validation set that consists of a small number of annotated data of case B. Competition participating teams  develop AI/ML/DL models for Covid-19 prediction. Performance of the different approaches will be evaluated on a test set (that contains data of case B) in terms of the ‘macro’ F1 score.

\section{The COV19-CT-DB Database}

COV19-CT-DB \cite{b300},  which we have developed, contains  3-D chest CT scans, collected in various medical centers. The database includes 7,756 3-D CT scans; 1,661 are COVID-19 samples, whilst 6,095 refer to non COVID-19 ones. There are about 2,500,000 images included in these datasets.  All have been anonymized. 724,273 images refer to the COVID-19 class, whilst 1,775,727 slices belong to non COVID-19 class
 \cite{arsenos2022large}. 
 
Table \ref{covctdb} presents a summary of the main elements of COV19-CT-DB.

\begin{table}[b!]
\caption{COV19-CT-DB: main elements}
\label{covctdb}
\centering
\scalebox{.9}{
\begin{tabular}{|c|c|}
\hline
Elements & Values \\
\hline
\hline
number of 3-D CT scans    & \begin{tabular}{@{}c@{}} 1,661 COVID \\  6,095 non-COVID \end{tabular} \\
\hline
number of 2-D images    & \begin{tabular}{@{}c@{}} 724,273 COVID \\  1,775,727 non-COVID  \end{tabular} \\
\hline
number of images in scan series   & 50 - 700 \\ 
\hline
size of images & $512 \times 512$ \\
\hline
  \end{tabular} 
}
\end{table}

\begin{figure}[h!]
\centering
\includegraphics[height=3.6cm]{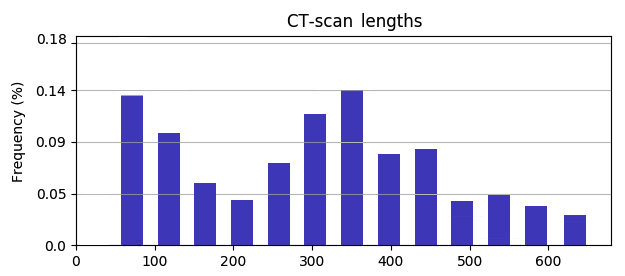}
\caption{COV19-CT-DB: 3-D scan length histogram }
\label{histograms}
\end{figure}

Figure \ref{histograms} analyzes the length of the CT scan series, presenting their histogram. This shows the differences regarding the length of 3-D CT scans in  COV19-CT-DB; these are caused by various reasons, including the requested resolution analysis, or the specific features of the used equipment. 

\begin{figure*}[h!]
\centering
\includegraphics[width = 0.19\linewidth]{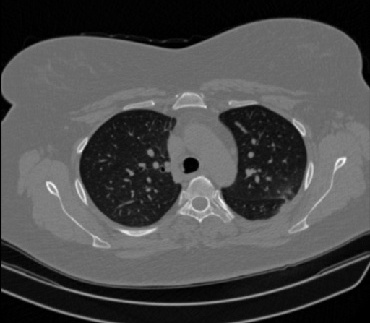}
\includegraphics[width = 0.19\linewidth,clip,trim={0 0.3cm 0 0}]{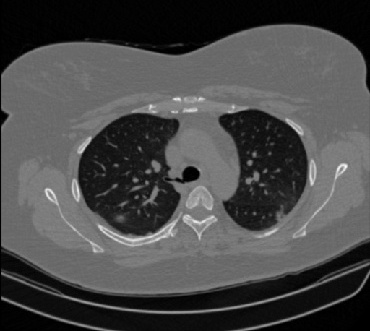}
\includegraphics[width = 0.19\linewidth]{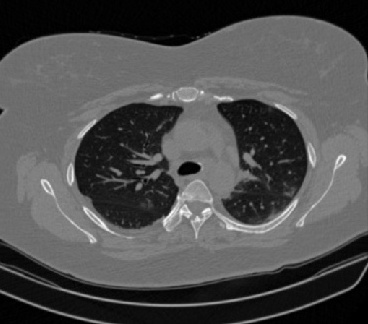}
\includegraphics[width = 0.19\linewidth]{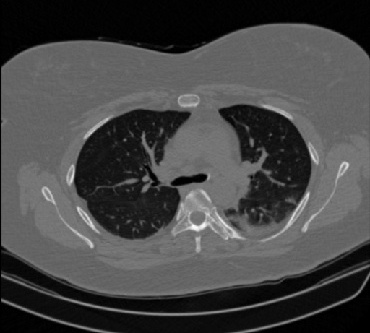}
\includegraphics[width = 0.19\linewidth]{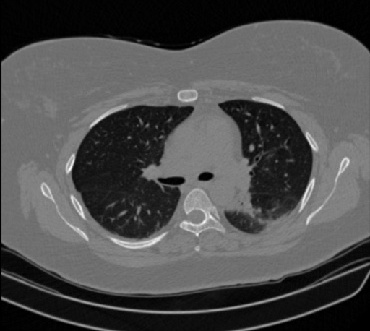} 
\\
\includegraphics[width = 0.19\linewidth]{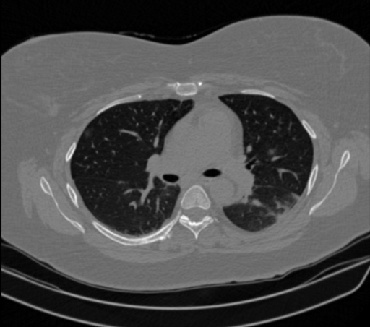}
\includegraphics[width = 0.19\linewidth,clip,trim={0 0.3cm 0 0}]{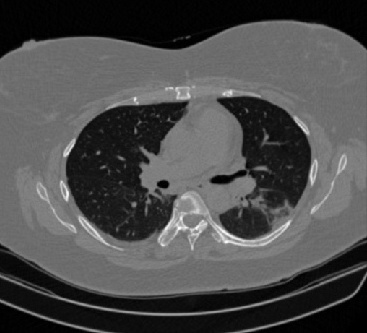}
\includegraphics[width = 0.19\linewidth]{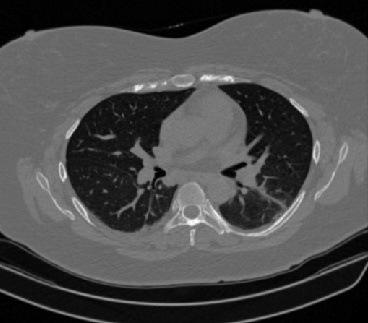}
\includegraphics[width = 0.19\linewidth]{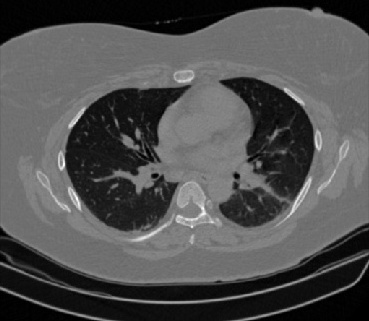}
\includegraphics[width = 0.19\linewidth]{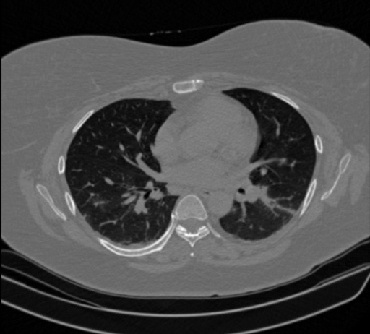}
\caption{Slices from a COVID-19 case in COV19-CT-DB} 
\label{covid_0}
\end{figure*}

\begin{figure*}[h!]
\centering
\includegraphics[width = 0.19\linewidth]{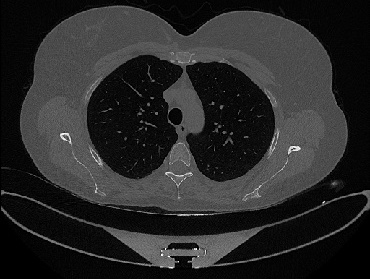}
\includegraphics[width = 0.19\linewidth]{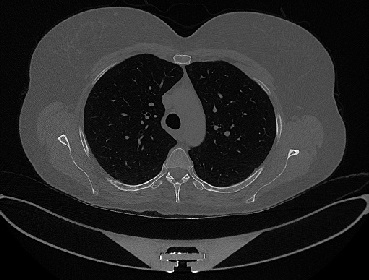}
\includegraphics[width = 0.19\linewidth]{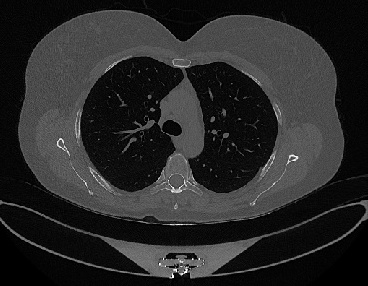}
\includegraphics[width = 0.19\linewidth]{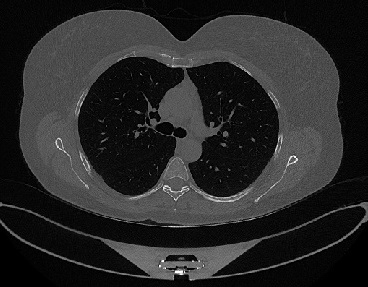}
\includegraphics[width = 0.19\linewidth]{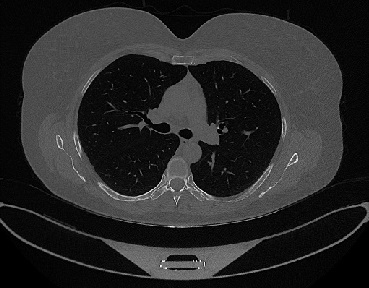}
\\
\includegraphics[width = 0.19\linewidth]{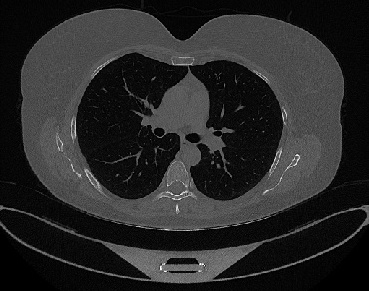}
\includegraphics[width = 0.19\linewidth]{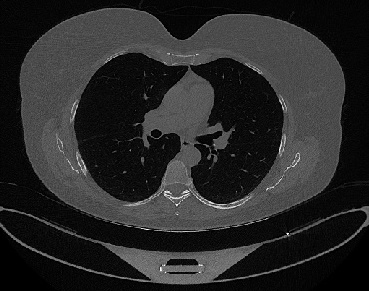}
\includegraphics[width = 0.19\linewidth]{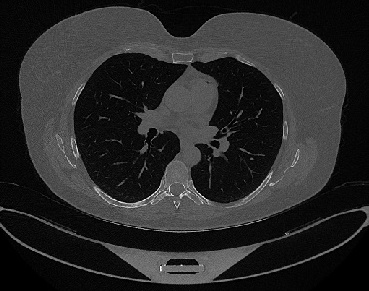}
\includegraphics[width = 0.19\linewidth]{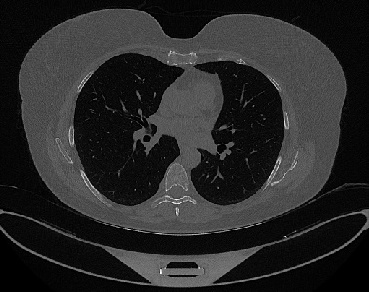}
\includegraphics[width = 0.19\linewidth]{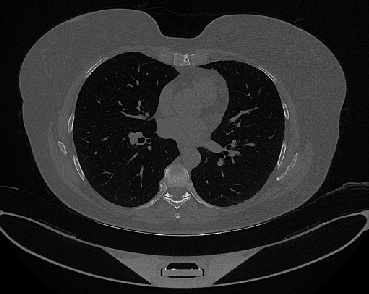}
\caption{Slices from  non COVID-19 case in COV19-CT-DB} 
\label{non_covid_1}
\end{figure*}

It should be mentioned that for explainability purposes \cite{Tailor, kollias2020transparent, kollia2019predicting}, an anchor set was generated for the COV19-CT-DB database \cite{b300}.  This included 11 anchors, each representing a respective 3-D CT scan obtained through an appropriate clustering procedure. 
Figure \ref{covid_0} shows a series of slices from a COVID-19 case, whereas Figure \ref{non_covid_1} shows a series of slices from a non COVID-19 case.

The first Challenge on COVID-19 detection is based on extract of this database. The training set contains, in total, 1358 3-D CT scans.  The validation set consists of 326 3-D CT scans.  The number of  COVID-19 and of  Non-COVID-19 cases in each set are shown in Table \ref{splits_ch1}. 

\begin{table}[h]
\caption{Data samples in each Set in Covid-19 Detection Challenge}
\label{splits_ch1}
\centering
\scalebox{1.}{
\begin{tabular}{| c | c| c |}

\hline Set & Training & Validation\\
\hline
\hline
COVID-19   & 703 & 170 \\
\hline
Non-COVID-19 & 655 & 156 \\
\hline
\end{tabular}
}
\end{table}

The second Challenge on COVID-19 Domain Adaptation is also based on extract from this database. The CT scans utilized have been sourced from a variety of hospitals and medical centers, providing a diverse range of data for analysis. The dataset has been partitioned into distinct training, validation and test subsets. 

239 3-D CT scans have been annotated and provided as training set to the participants, with 178 3-D CT scans constituting the validation set. In addition, 494 3-D CT scans have been provided without annotations, as shown in \ref{splits_ch2}  so that they can be used by the participants in the adaptation process. 

\begin{table}[h]
\caption{Data samples in each Set in Covid-
19 Domain Adaptation Challenge}
\label{splits_ch2}
\centering
\scalebox{1.}{
\begin{tabular}{| c | c| c |}

\hline Set & Training & Validation\\
\hline
\hline
COVID-19   & 120 & 65 \\
\hline
Non-COVID-19 & 119 & 113 \\
\hline
Non-annotated & 494 & - \\
\hline
\end{tabular}
}
\end{table}


\section{The baseline configurations}

\subsection{COVID-19 detection \& domain adaptation baselines}


The baseline architecture adopted for both Challenges, namely the COVID-19 Detection Challenge and the Covid-19 Domain Adaptation Challenge, is a CNN-RNN architecture \cite{b300, arsenos2023data,cis,kollias2023btdnet}.

The input 3-D CT scans have been padded to achieve a uniform length $t$, ensuring that every 3-D CT scan contains $t$ slices. The entire unsegmented sequence \cite{salpea2023medical} of 2-D slices from a CT scan is then fed into the CNN component. This CNN component conducts localized analysis on a per-2D-slice basis, primarily extracting features from the lung regions. The objective is to facilitate diagnosis using the entire 3-D series of CT scans, mirroring the annotations provided by medical experts.

Subsequently, the RNN component analyzes the CNN features of the complete 3-D CT scan, sequentially traversing from slice $0$ to slice $t-1$. The outputs of the RNN component are forwarded to a Fully Connected layer and subsequently to an output layer utilizing a softmax activation function to provide the COVID-19 diagnosis. We also include a Dropout layer before the Fully Connected one.

In the second Challenge (Covid-19 Domain Adaptation), we employed Monte Carlo Dropout to assess uncertainty while training the CNN-RNN architecture using data from both case A (annotated) and case B (annotated). Monte Carlo Dropout is a technique that involves performing multiple forward passes through the network with dropout activated during inference, allowing us to capture the model's inherent uncertainty. Subsequently, we annotated the non-annotated data from case B based on the model's predictions, specifically considering COVID instances where the model exhibited a high confidence level. This approach enabled us to leverage the model's uncertainty estimates to adapt to the non-annotated data of case B.





\subsection{Pre-Processing \&  Implementation Details}

In the pre-processing stage, all 2-D CT slices have been extracted from respective DICOM images. Next,  voxel intensity values were computed through a window of $350$ Hounsfield units (HU)/$-1150$ HU; they were then normalized in the range $[0, 1]$. 
Data augmentation was also performed, including random rotation in [-10\degree, 10\degree] and horizontal flip \cite{zheng2020deep,huang2020serial} to extract region of interests, such as lung areas in the 2-D images.

As far as implementation of the  baseline approach is concerned, the following models have been used: i) we adopted the CNN ResNet50 model; on top of it we included a global average pooling,  as well as a batch normalization layer and dropout (with keep probability 0.8), ii) we used a single one-directional GRU RNN layer comprising 128 neurons. The model input consisted of the 3-D CT scans. Each 2-D image was resized from its  size of $512 \times 512 \times 3$ to  $224 \times 224 \times 3$. We selected a confidence threshold of 70\% to determine high-confidence annotations for non-annotated data in the Domain Adaptation Challenge.

Batch size was equal to 5 (i.e, at each iteration our model processed 5 CT scans) and the input length 't' was 700 (the maximum number of slices found across all CT scans). We utilized the softmax cross entropy as loss function for training both baseline methods. Adam optimizer was used with learning rate   $10^{-4}$. Training was performed on a Tesla V100 32GB GPU.

\section{Experimental Results}

This section describes a set of experiments evaluating the performance  of the baseline configurations.

Table \ref{3dcnn_rnn} shows the performance of the network over the validation sets in both Challenges, after training with the training datasets, taking into account that there exists only a single label for the whole CT scan and no labels for each CT scan slice \cite{b300}. 

In both Challenges , the performance of the baseline methods were evaluated in terms of the macro F1 score. The macro F1 score is defined as the unweighted average of the class-wise/label-wise F1-scores, i.e.,  the unweighted average of the COVID-19 class F1 score and of the non-COVID-19 class F1 score.

\begin{table}[t]
\caption{Performance of baseline model in each Challenge}
\label{3dcnn_rnn}
\centering
\scalebox{1.}{
\begin{tabular}{|c|c|}
\hline
 Challenge  &  \multicolumn{1}{c|}{'macro' F1 Score}\\
 \hline
 \hline
 COVID-19 Detection   &   0.78  \\
\hline
 COVID-19 Domain Adaptation   &   0.73  \\
\hline
\end{tabular}
}
\end{table}



\section{Conclusions and Future Work}

In this paper we present  the 4th COV19D Competition and particularly the two Challenges that it contains: the first on COVID-19 detection and the second on COVID-19 domain adaptation.  We  provide a short description of the COV19-CT-DB, extracts from which are used in the two Challenges. We also present the baseline approaches and their performance in the Challenges. 



\bibliographystyle{IEEE}
\bibliography{egbib}

\end{document}